\documentclass[letterpaper,twocolumn,twocolappendix,appendixfloats]{emulateapj}
\usepackage{float}
\usepackage{epsfig}
\usepackage{graphicx}
\usepackage{graphics}
\usepackage[latin1]{inputenc}
\usepackage{latexsym}
\usepackage{rotfloat}

\shorttitle{Variables from TNTS}
\shortauthors{Yao et al.}

\begin{document}

\title{Photometry of Variable Stars from THU-NAOC Transient Survey I: the first 2 years }

\author{Xinyu Yao\altaffilmark{1,2,3}, Lingzhi Wang,\altaffilmark{4,3,5}, Xiaofeng Wang\altaffilmark{3}, Tianmeng Zhang\altaffilmark{4}, Juncheng Chen\altaffilmark{3}, \\ Wenlong Yuan\altaffilmark{6,3}, Jun Mo\altaffilmark{3}, Wenxiong Li\altaffilmark{3}, Zhiping Jin\altaffilmark{2}, Xuefeng Wu\altaffilmark{2}, \\ JunDan Nie\altaffilmark{4}, Xu Zhou\altaffilmark{4}}

\altaffiltext{1}{Qinghai Normal University, Xining 810008,China}

\altaffiltext{2}{Purple Mountain Observatory, Chinese Academy of Sciences, Nanjing, 210008, People's Republic of China. xfwu@pmo.ac.cn; jin@pmo.ac.cn
}

\altaffiltext{3}{Physics Department and Tsinghua Center for Astrophysics (THCA), Tsinghua University, Beijing
100084, China. wang\underline{ }xf@mail.tsinghua.edu.cn
}

\altaffiltext{4}{Key Laboratory of Optical Astronomy, National Astronomical Observatories, Chinese Academy of Sciences, Beijing 100012, China. wanglingzhi@bao.ac.cn
}

\altaffiltext{5}{Chinese Academy of Sciences South America Center for Astronomy, Camino El Observatorio 1515, Las Condes, Santiago, Chile
}

\altaffiltext{6}{George P. and Cynthia Woods Mitchell Institute for Fundamental Physics and Astronomy, Department of Physics \& Astronomy, Texas A\&M University, 4242 TAMU, College Station, TX 77843-4242, USA
}

\begin{abstract}

In this paper, we report the detections of stellar variabilities from the first 2-year observations of sky area of about 1300 square degrees from the Tsinghua University-NAOC Transient Survey (TNTS). A total of 1237 variable stars (including 299 new ones) were detected with brightness $<$ 18.0 mag and magnitude variation $\gtrsim$ 0.1 mag on a timescale from a few hours to few hundred days. Among such detections, we tentatively identified 661 RR Lyrae stars, 431 binaries, 72 Semiregular pulsators, 29 Mira stars, 11 slow irregular variables, 11 RS Canum Venaticorum stars, 7 Gamma Doradus stars, 5 long period variables, 3 W Virginis stars, 3 Delta Scuti stars, 2 Anomalous Cepheids, 1 Cepheid, and 1 nove-like star based on their time-series variability index $J_s$ and their phased diagrams. Moreover, we found that 14 RR Lyrae stars show the Blazhko effect and 67 contact eclipsing binaries exhibit the O'Connell effect. Since the period and amplitude of light variations of RR Lyrae variables depend on their chemical compositions, their photometric observations can be used to investigate distribution of metallicity along the direction perpendicular to the Galactic disk. We find that the metallicity of RR Lyrae stars shows large scatter at regions closer to the Galactic plane (e.g., $-$3.0 $<$ [Fe/H] $<$ 0) but tends to converge at [Fe/H]$\sim$ $-$1.7 at larger Galactic latitudes. This variation may be related to that the RRAB Lyrae stars in the Galactic halo come from globular clusters with different metallicity and vertical distances, i.e. OoI and OoII populations, favoring for the dual-halo model.

\end{abstract}


\section{Introduction}
Time-Domain Astronomy (TDA) has emerged as a key field of current astronomy and astrophysics \citep[]{TDA}. This offers new routes to astrophysical understanding of stellar and galactic activities such as variable stars, novae, supernovae, gamma-ray bursts, AGN and quasars etc.. The Tsinghua University-NAOC Transient Survey (TNTS) is a four-year project designed to search such transients (Zhang et al. 2015). The observations were carried out in a clear filter with a 0.6-m Schmidt telescope located at Xinglong station of NAOC. Under a typical weather condition at Xinglong station (e.g., with a seeing of about 2 arcsec), this telescope and the CCD system can reach a detection limit of about 19.5 mag (3$\sigma$) in the clear filter for an exposure of 60$s$. The relatively short cadence (e.g., 3-4 days) and broad synoptic coverage of the sky areas provide an excellent chance to search for and study variable stars of different time scales. In this paper, we present the result of searching for variable stars from these survey data.

This work presents the result of searching for variable stars from the TNTS data accumulated during the period from Oct. 2012 to Aug. 2014. The observations and data reductions are described in Section 2. The time-series photometry from TNTS images is presented in Section 3. The classifications of variable stars and analysis are given in Section 4. We summarize in Section 5.

\section{Observations and Data Reductions}
The TNTS is conducted with a 0.6-m Schmidt telescope equipped with a 4k$\times$4k CCD. The CCD has a pixel scale of 1.3 arcsec pixel$^{-1}$ and a field of view (FoV) of 1.5 degrees $\times$ 1.5 degrees. It's designed to operate for four years starting from October 2012. This survey covers a sky area of $\sim$1800 square degrees with Galactic latitude $|b|$ $>$ $10^\circ$ and $0^\circ < \delta < 60^\circ$. It usually takes about 2 minutes to get an image for a specific sky zone, including 60-s exposure for the selected target,
22-$s$ readout time, and 30-$s$ movement and stabilization for the telescope. In order to efficiently rule out the cosmic rays and moving objects, we took two exposures for each sky zone with a temporal interval of 1-1.5 hours and the whole survey area can be visited every 3 to 4 days (see details in Zhang et al. 2015).

About 85,000 images were accumulated from our survey performed during the period from Oct. 2012 to Aug. 2014, which cover about 1800 square degrees of the northern sky. In this study, however, we only chose the fields with repeated observations $>$ 40 times to detect variable stars. This selection criterion results in that the observations of a sky area of $\sim$ 1300 square degrees can be used for such an analysis. The TNTS images were reduced using the IRAF \footnote{IRAF, the Image Reduction and Analysis Facility, is distributed by the National Optical Astronomy Observatory, which is operated by the Association of Universities for Research in Astronomy(AURA), Inc. under cooperative agreement with the National Science Foundation(NSF). http://iraf.noao.edu/} standard procedure, including the corrections for bias, flat field, and removal of cosmic rays. The positions of those images were then registered to each other using the Guide Star Catalogue (GSC) \citep[]{GSC}, with an accuracy of 0.5 arcsec \citep[]{Zhou2003}. The SExtractor \citep[]{Bertin1996} was used to perform aperture photometry on the chosen images. The radius of aperture photometry was set as 5 pixels (6.8 arcsec), with a sky annulus size of 24 pixels (31.2 arcsec).

\begin{figure}[t]
\begin{center}
\includegraphics[width=0.45\textwidth]{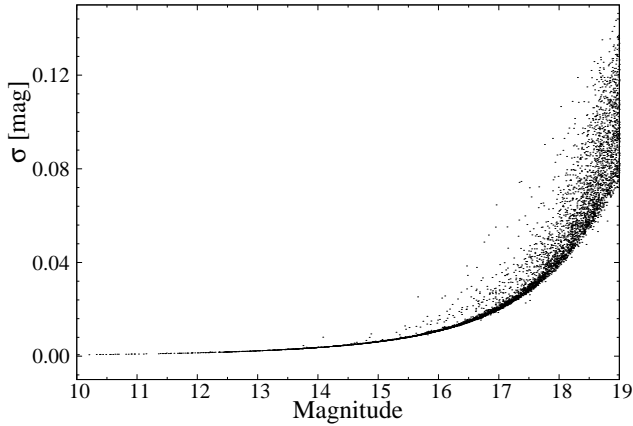}
\caption{The photometric error versus magnitude in a clear band for a typical image of the field SE65 from the TNTS.}
\label{fig:err}
\end{center}
\end{figure}

Since the stars on the TNTS images are not very crowded, we selected the parameter "BEST OF MAG" of SExtractor as output results. The detection threshold was set as 2$\sigma$ above the sky background. To alleviate the effect of nonuniform background on the photometry, i.e., the moonlight and clouds etc., we set the BACKGROUND TYPE as LOCAL. A range of 5000-17000 stars can be detected on the TNTS images. In our study we take the typical sky field SE65 as an example for illustration of photometric error. For stars brighter than 17.0 mag, the photometric error is generally less than 0.02 mag according to the output result from the SExtractor (see Fig. \ref{fig:err}).

\section{Time-series photometry}
To get the time-series photometry of each star, we feed the matching tool SRPMatch with photometry catalogs produced by
SExtractor for two images of the sky field. SRPMatch is a command line tool embedded in the Swift Reduction Package
(hereafter SRP \footnote{ https://pypi.python.org/pypi/SRPAstro}), which can be used to match photometry catalogues. We set the match tolerance as 1.5 arcsec ($\sim$ 1 pixel). For each field, one is the reference image with the maximum number of detected stars and the other is the random image from the same field. We calibrated our time-series photometry by the following two steps: magnitude calibration and rescaling of the photometric errors.

\subsection{Magnitude Calibration}

To calibrate magnitudes of images taken at different epochs, it is necessary to calculate the magnitude offset of each image relative to the reference image. Fig. 2 shows the magnitude offsets measured for the sky field SE65 (which is the same as that used in Fig. \ref{fig:err}) and its reference frame. One can see that the photometric error progressively increases for stars fainter than 14.0 mag.
On the other hand, the stars brighter than 12.5 mag tend to be at the edge of saturation on the TNTS images. We thus chose those stars with magnitudes between 12.5 mag and 14.0 mag to determine the magnitude offset for each image. To mitigate the effect of nonuniform background due to moonlight or obscuration by clouds, we divided every image into 16 small subsections and then calculated the offset of each subsection to perform the photometric calibrations. Outlier rejections of 3$\sigma$ were applied iteratively to make sure that no variables were involved in the calculations. The median value of the magnitude offsets of 16 subsections is almost the same as that of the whole image, e.g., $\Delta M_{zp}=0.010\pm0.001$ for the sky field SE65. The uncertainty in the magnitude offset is so small (relative to the photometric errors from SExtractor) that we neglected it in the following calculations.

\begin{figure}[h]
\begin{center}
\includegraphics[width=0.45\textwidth]{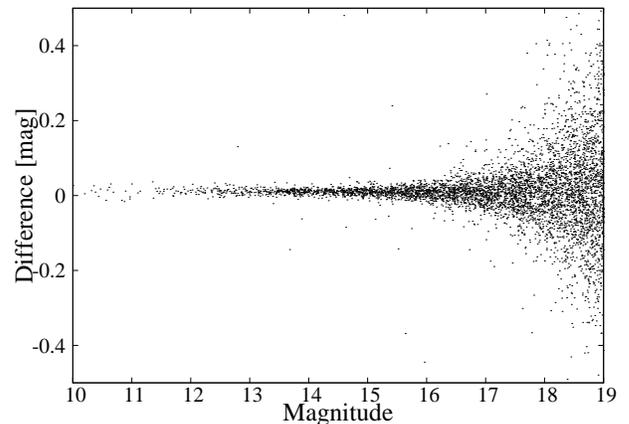}
\caption{Magnitude offsets between the same image of the field SE65 used in Fig. \ref{fig:err} and the reference frame of the same field.}
\label{fig:zpt}
\end{center}
\end{figure}

Previous studies indicate that unfiltered magnitude is approximately equivalent to that in the R band of PPMX system
\citep[Position and Proper Motions eXtended,][]{PPMX}, with an uncertainty of 0.2-0.3 mag, depending on the number of
flux-calibrated stars available in the field (e.g., Li et al. 2003, 2011). For each field of the TNTS, we match the reference image with the PPMX catalogue to choose appropriate stars to calibrate the unfiltered magnitudes. Owing to the large FoV of the TNTS images, the number of stars that can be used for the flux calibration is relatively large. The median uncertainty in the photometric calibration
for the 600 fields used in our study is $\sim$0.010$\pm$0.005 mag, which is much smaller than that from \citet{Li03}.

\subsection{Rescaling the Measured Photometric Errors}

\begin{figure}[ht]
\begin{center}
\includegraphics[width=0.45\textwidth]{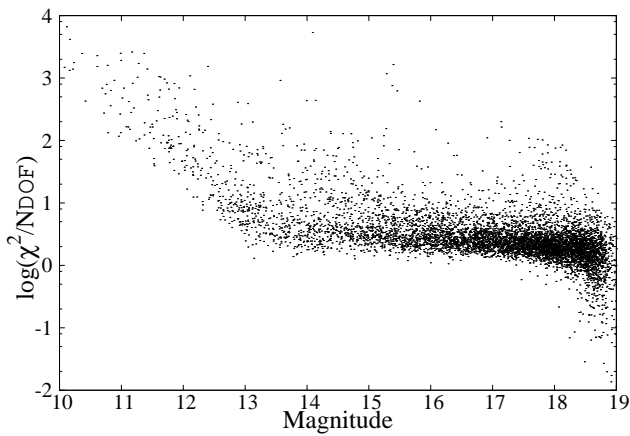}
\caption{The $Log (\chi^2/N_{DOF})$ distribution as a function of the weighted mean magnitude for stars in the field SE65 before rescaling their photometric errors.}
\label{fig:chi1}
\end{center}
\end{figure}

For some reason, the photometric errors given by SExtractor might be overestimated or underestimated. We would either miss real variables, or select bogus variables as real ones \citep[]{Chi2}. Assuming that the photometric errors roughly follow the Gaussian distribution, the logarithm of the $\chi^2/N_{DOF}$ of stars with constant luminosities should be close to zero \citep[see the right panel of Fig. 6 from ][]{Chi2}. The $\chi^2/N_{DOF}$ is defined as
\begin{equation}
\chi^2=(\frac{m-\bar{m}}{\sigma_{m}})^{2}
\end{equation}

\begin{equation}
N_{DOF}=N-M.
\end{equation}
Where $\bar{m}$ is the weighted mean value (the weighting factor is the inverse square of the error $\sigma_{m}$) of the time-series magnitude $m$ of a star, $N$ is the number of measurements, and $M$ is the number of parameters (here we let it be one). Take the field SE65 as an example, Fig. \ref{fig:chi1} shows the plot of the logarithm of the $\chi^2/N_{DOF}$ against weighted mean magnitude for all of the stars. One can see that the brightest stars ($\sim12.0$ mag) have $\chi^2/N_{DOF}\sim100$, and their photometric errors are thus
seriously underestimated. The possible reasons for this discrepancy includes underestimation of the flat fielding error and imperfect aperture photometry etc. To eliminate this inconsistency, the magnitude error $\sigma$ is multiplied by an appropriate scaling factor F, following the method proposed by \citet[]{Chi2}.

\begin{figure}[ht]
\begin{center}
\includegraphics[width=0.45\textwidth]{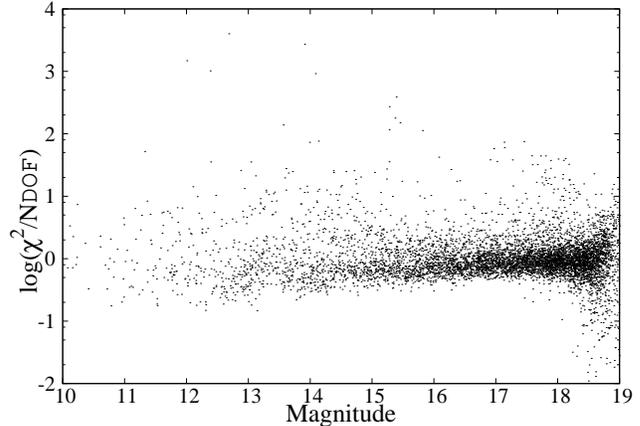}
\caption{The same diagram as the Fig. \ref{fig:chi1} but after rescaling the photometric errors.}
\label{fig:chi2}
\end{center}
\end{figure}

In detail, we first binned the data $\chi^2/N_{DOF}$ at an interval of 0.1 magnitude. In the logarithm plot, we used {\bf a second-order polynomial} to fit the data for stars with brightness $<$ 13.5 mag. For stars with brightness between 13.5 mag and 18.0 mag, we used a linear function to fit. While those stars with weighted mean magnitudes fainter than 18.0 mag were discarded in the fit due to relatively larger photometric errors. We then applied the scaling factor $F$ to rescale the photometric error of each star. The distribution of the rescaled photometric errors for stars in the field SE65 is shown in Fig. \ref{fig:chi2}.

\ \par

\section{Variables from the TNTS}

\subsection{Searching for Variable Stars}

The variable candidates were initially selected using Welch-Stetson variability index $J_s$ \citep{Stetson}. We used VARTOOLS \citep[]{vartools} to compute the $J_s$ index based on the rescaled errors of photometry. For each star one can calculate the variability index $J$ as
\begin{equation}
J=\frac{\sum_{k=1}^{n} w_k \mathrm{sgn}(P_k)\sqrt{|{P_k}|}}{\sum_{k=1}^{n} w_k},
\label{eq:stetj}
\end{equation}
where $\mathrm{sgn}$() is the sign function. We considered $n$ pairs of observations, and each has a weight $w_k$,
\begin{equation}
P_k=\left\{ \begin{array}{ll}
        \delta_{i(k)}\delta_{j(k)}, 	& \mbox{if $i(k)\neq j(k)$}, \\
	\delta_{i(k)}^2 -1, 		& \mbox{if $i(k)=j(k)$}
\end{array} \right.
\end{equation}
where $P_k$ is the product of the normalized residuals of the two paired observations $i$ and $j$, and
\begin{equation}
\delta=\sqrt{\frac{n}{n-1}}\frac{m-\bar{m}}{\sigma_{m}}
\label{eq:delta}
\end{equation}
represents the magnitude residual from the mean value scaled by the standard error \citep[]{Stetson}. In detail, $n$ is the
total number of observations contributing to the mean. The final variability index is
\begin{equation}
J_{s}=J\frac{\sum w_k}{w_{max}}.
\label{eq:stetj1}
\end{equation}
In our calculations, we considered the two observations taken on the same night as a pair and give each of them a weight of 0.5. Otherwise a weight of 1.0 was given if there is only a single observation.

\begin{figure}[ht]
\begin{center}
\includegraphics[width=0.45\textwidth]{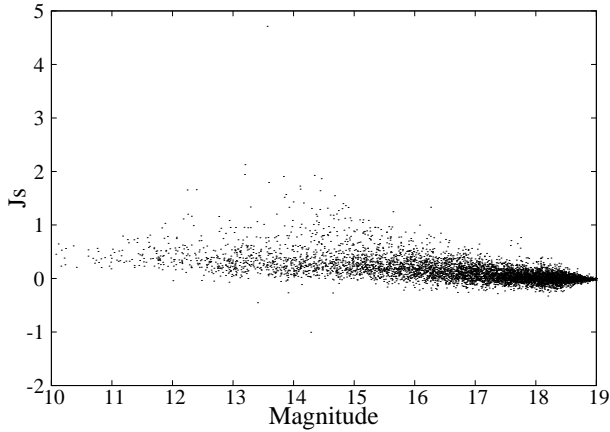}
\caption{Distribution of the $J_s$ index calculated for stars in the sky field SE65.}
\label{fig:Js}
\end{center}
\end{figure}

As an example, we show in Fig. \ref{fig:Js} the distribution of $J_s$ obtained for stars in the field SE65. As it is expected most stars have $J_s$ values close to zero. The mean $J_s$ value of 600 sky fields involved in this study is 0.07. For each field, we chose those stars with $J_s$ larger than the mean value by 3$\sigma$ as candidates of variable stars.

We searched for periodic variabilities among candidates using Lomb-Scargle (L-S) method \citep[]{L, S, Press1989, Press1992} and Phase Dispersion Minimization (PDM) method \citep[]{PDM}. The L-S method applies the statistical properties of least-square frequency analysis of unequally spaced data on a series of test periods. We set test periods ranging from 0.1d to 600d with a bin size of 0.1d. We determined the most likely period for each star in terms of the power value given by the L-S method. Then we performed PDM method to double check the periods detected from the L-S method since the PDM method is useful for data sets with gaps, non-sinusoidal variations, poor time coverage or other problems that would make Fourier technique unusable. The folded light curves were then established using the period from the L-S/PDM or twice of that for a binary if necessary. The period derived from the PDM method was only adopted if the phased light curve folded by the PDM has smaller scatter than that from the LS method. We examined the light curve and phase diagram for each candidate to pick out variables and throw out those showing alias variations (e.g., one day, half day, one third day and so on) and having large dispersions. Finally we found 1237 variables with magnitude variations $\gtrsim$ 0.1 mag and brightness $<$ 18.0 mag based on the L-S method.

\begin{center}
\begin{figure}[th]
\includegraphics[width=0.45\textwidth]{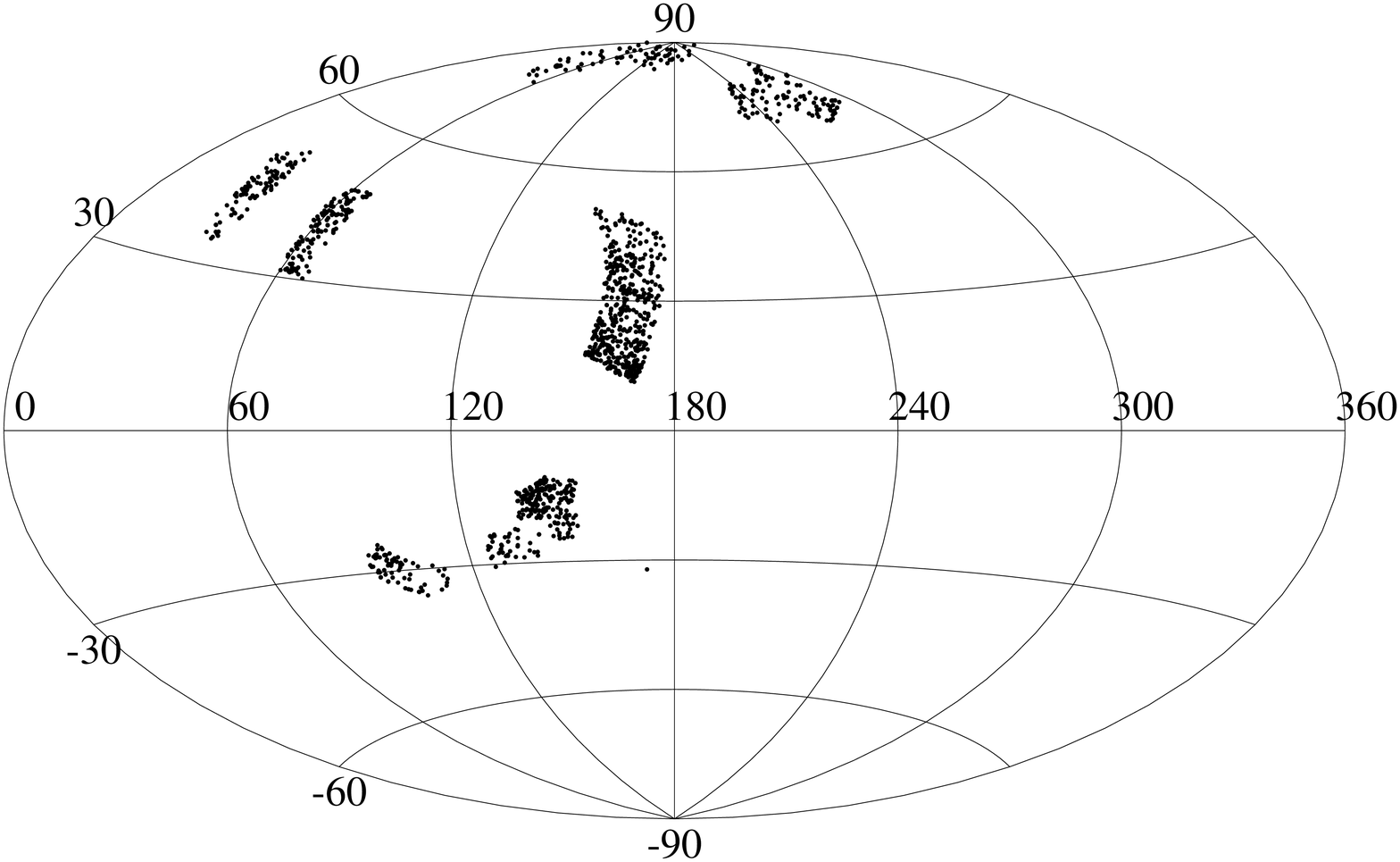}
\caption{Distribution of the TNTS variable stars in the Galactic coordinates. }
\label{fig:gal}
\end{figure}
\end{center}

All of these candidates had been cross-checked with those listed in the VSX database \citep[the International Variable Star Index,][]{Watson}, which is created by amateur astronomers of American Association of Variable Star Observers (AAVSO) \footnote{AAVSO: http://www.aavso.org/vsx/}. We found that 938 out of these 1237 objects have been included in this database with 438 ones from the Catalina Sky Survey (CSS) sample \citep[]{CSS} and 106 ones from the Lincoln Near-Earth Asteroid Research (LINEAR) survey sample \citep[]{Palaversa13}, and the rest 299 ones can be considered as new variables. Fig. \ref{fig:gal} shows the distribution of these variables, in which they are displayed according to the Galactic coordinates. The parameters of the first 20 ones of variable stars from variable star candidates table are listed in Table \ref{tab:var}. Column 1 and 2 list the survey field and the ID number of the field; Column 3 and 4 give the right ascension and declination of the relevant field; Column 5 gives the mean unfiltered magnitude (calibrated with the R-band magnitudes of PPMX catalogue); Column 6 gives the line-of-sight extinction based on Schlafly \& Finkbeiner(2011); Column 7 gives the amplitude of variation (magnitude from maximum to minimum); Column 8 gives the variability index; Column 9 and 10 give the periods of the variables from our measurements and the corresponding method; Column 11 gives the period from the VSX; Column 12 gives the minimum light of the variables detected in the observation season; Column 13 gives a tentative classification of the variable star whenever possible; Column 14 shows the additional information about the variable stars, including previous identifications from the VSX database (e.g., CSS or LINEAR). Note that, for those known variables, the periods estimated from the TNTS are highly consistent with those given by the VSX database except about thirty ones (see Figure 7). The phased light curves of new periodic variable stars are shown in the appendix.

\begin{center}
\begin{figure}[th]
\includegraphics[width=0.45\textwidth]{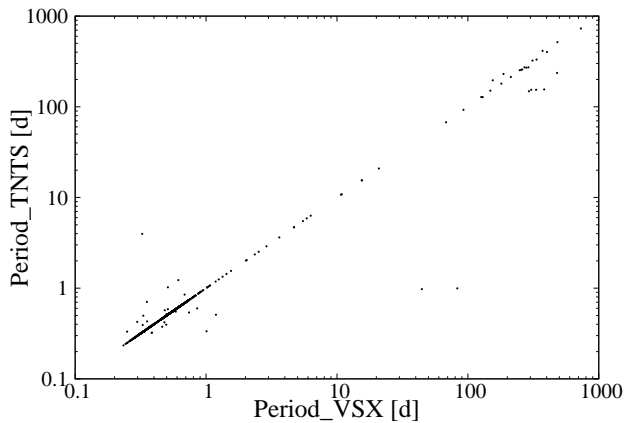}
\caption{The measured periods from the TNTS data versus those from the AAVSO VSX for the variables in common.}
\label{fig:per}
\end{figure}
\end{center}

\subsection{Classification of Variables}

Most of the variables exhibit regular variations in periods and amplitudes, i.e., RR Lyraes and eclipsing binaries. Table \ref{tab:typ} lists the types of the 1237 variables from the first 2-year survey data, which are classified based on the shapes of their light curves, amplitudes and periods of phase diagrams. Half of the detected variables are found to be RR Lyres according to their specific periods and light curve shapes. The remained variables include 431 eclipsing binaries, 72 Semiregular pulsators, 29 Mira stars, 11 slow irregular variables, 11 RS Canum Venaticorum stars, 7 Gamma Doradus stars, 5 long period variables, 3 W Virginis stars, 3 Delta Scuti stars, 2 Anomalous Cepheids, 1 Cepheid, and 1 nove-like star. Note that our tentative classifications of variable stars are based only on the unfiltered photometry.

\addtocounter{table}{1}
\begin{deluxetable}{lrrc}
\tablenum{2}
\tablecaption{Distribution of Types of Variables from the TNTS}
\label{tab:typ}
\tablehead{\colhead{Variable Type} & \colhead{N} & \colhead{\%} &  \colhead{N(New)}}
\startdata
RR lyr              &  661  & 53.44  &  59 \\
Eclipsing binaries  &  431  & 34.84  & 157 \\
Semiregular pulsators&  72 &  5.82 & 68 \\
Mira (LPV)          &   34  & 2.75  & 0 \\
Irregular           &  11  &  0.89  &  0 \\
RS Canum Venaticorum &  11  &  0.89 &  2   \\
Gamma Doradus stars &   7  &  0.57  & 7 \\
W Virginis          &  3  &  0.24 & 0 \\
Delta Scuti Stars   &  3  &  0.24  &  3\\
Anomalous Cepheids  &   2  &   0.16   &   1  \\
Cepheids             &  1  &  0.08   &   1  \\
Nova                &   1 &  0.08   &   1
\enddata
\end{deluxetable}

\ \par

\subsubsection{RR Lyrae Stars}

RR Lyrae stars, located at intersections of the instability strip and horizontal branch, belong to a subclass of pulsating variable stars. They are old, low-mass, and metal-poor "Population II" stars which are usually found in globular clusters. According to their pulsation modes, the RR Lyrae stars can be divided into three subtypes: RRAB-type (pulsating in the fundamental mode); RRC-type (pulsating in first overtone); and RRD-type \citep[pulsating simultaneously in two modes,][]{Smith1995}. Meanwhile, different subtypes of variable stars have different periods and amplitudes. In our sample, there are 526 RRAB-type stars, 126 RRC-type stars and 9 RRD-type stars in light of the periods and amplitudes of variations, which are shown in Figure 8. For our sample, the mean period of the RRAB-type variables is 0.57 d, while the RRC-type ones have the mean period of 0.34 d. The mean amplitude of the light variation for the RRAB subgroup (0.84 mag) is relatively larger than that of the RRC subgroup (0.44 mag).

The light curves of RRAB Lyrae stars are characterized by variations with a fast rising followed by a slow declining \citep{Jurcsik2009a,Cacciari05}. The variable star S100-3820 (see Fig. A10) was originally detected by AASVO VSX and identified as a RRAB-type star. However, our observation shows that it has a period of $\sim$0.5 d and a variation amplitude of 0.37 mag, lying at the boundary between RRAB-type and RRC-type (see Figure 8). According to our result, its rise time is equal to the decline time, which likely just evolved off the zero-age horizontal branch (ZAHB) and exhibits light variation with a fixed amplitude and a longer period \citep[][\S3.2]{Cacciari05}. Another two interesting RR Lyrae stars in our sample are S037-10947 (see Fig. A10) and S585-9662 (see Fig. A5), which lie at a similar position as S100-3820 in Figure 8. Although they have typical features of this type of variable stars, i.e., fast rising and slow declining, they show a little bump close to the minimum light. More data are needed to determine the properties of these two interesting RR Lyrae stars.

Some RR Lyrae stars are known to have ``Blazhko effect'' which exhibits modulations in their periods, pulsation amplitudes, light curve shapes, and radial velocities \citep{Blazhko1907, Buchler11}. Previous surveys have found that 10\%-30\% of the RRAB-type variables have such a effect \citep{Alcock2003} and this ratio can reach up to 50\% according to a recent study by \citet{Jurcsik2009a}. Fourteen RRAB-type Lyrae stars of our sample are found to have phase modulations and can be tentatively classified as RRAB stars with the Blazhko effect. Obviously, we did not detect such a high ratio from our sample when compared with previous surveys. It is likely the current TNTS data may not be qualified enough to see most of the phase modulations. Increasing the cadence of the survey may help to discover more RR Lyrae stars with Blazhko effect.

The period-amplitude diagram for RR Lyrae stars, also termed as Bailey diagram, was used to investigate their properties in globular clusters more than a century ago (Bailey 1902). Inspecting our RRAB stars shown in Figure 8, we find that these variables have different locations in this diagram. For example the periods could differ by $\sim$0.1-0.2 day at given amplitude. There has been debate on whether Oosterhoff type or metallicity of the clusters accounts for these differences. The Oosterhoff-type classification of the Galactic globular clusters was based on the frequency distribution of their RR Lyrae variables' periods (Oosterhoff 1939). For example, RRAB stars in Oosterhoff I (OoI) clusters have mean periods of $\sim$0.55 day, while those in Oosterhoff II (OoII) clusters have mean periods of $\sim$0.64 day (i.e., Smith et al. 1995, Clement et al. 2001, Catelan 2009). On the other hand, the metallicity has been found to correlate with both
period and amplitude for the RR Lyrae stars (i.e., Marconi et al. 2015 and references therein). It has been found that the observed
period of RR Lyrae stars tend to become shorter with increasing metallicity (i.e., Preston 1959; Kunder \& Chaboyer 2009;
Fiorentino et al. 2015), while the pulsation amplitude decreases as the metallicity increases (i.e., Bono et al. 1996). Such an
amplitude-period-metallicity relation can be used to estimate metallicity for different RRAB Lyrae stars (Sandage 1982, Alcock et al. 2000,
Sandage et al. 2004, Kinemuchi et al. 2006). The metallicity dependence of the period-amplitude relation can be interpreted as an
opacity effect in stellar evolution (i.e., Bono et al. 1996), but a detailed explanation is beyond the scope of this paper. Note that
the effect of Oosterhoff classification on the period-amplitude relation may be in part coupled with that of metallicity, as OoI
clusters tend to have higher metallicity than clusters of OoII (Catelan 2009).


Note that the amplitudes of RR Lyraes shown in Figure 8 are measured in the unfiltered light curves, which are found to be systematically smaller by 0.15 mag than those measured in the $V$ band (i.e., Zorotovic et al. 2010). A similar offset was noticed in the unfiltered data from other surveys such as the CSS (Drake et al. 2013) and SSS (the Siding Spring Survey; Torrealba et al. 2015). We thus convert the unfiltered amplitude from the TNTS to the $V$-band value by adding an offset of 0.15 mag for each star of our sample. The corrected period-amplitude diagram of the RRAB-type stars is shown in the inset of Figure 8, where we add the reference line of OoI (solid line) based on eq. (11) in \citet[]{OoI/OoII}. Overplotted is the OoII line (dashed line) which follows the same trend as OoI but shifted in period by $\Delta$ $Log P$=+0.06. Following Miceli et al. (2008), a period shift of 0.05 d relative to the OoI line was adopted to divide our RRAB Lyrae stars into the OoI and OoII types. One can see that the majority of the TNTS RRAB Lyrae stars cluster around the OoI line, which is consistent with that inferred from the CSS sample extending up to $\sim$60 kpc above the galactic plane (see Fig. 5 of Drake et al 2013) and also consistent with that from the Siding Spring Survey (see Fig. 11 of Torrealba et al. 2015).

Since we have a relatively large sample of RRAB Lyrae stars locating from the disk to the halo of the Galaxy, we can carry on studies on the
spatial distributions of their metallicities and this helps trace history of the Galaxy formation (Catelan 2009; Pietrukowicz 2014).
The metallicity is estimated using the empirical period-amplitude-metallicity relation given by Sandage (2004), which
follows as:
\begin{eqnarray}
[Fe/H]=-(1.453\pm 0.027)A_V-(7.990\pm 0.091)logP{} \nonumber\\
       -(2.145\pm 0.025) \qquad \qquad \qquad \qquad \quad \quad \quad \
\label{eq:metallicity}
\end{eqnarray}
Where $A_V$ is the $V$-band amplitude corrected from the unfiltered value, $P$ represents the period. The amplitude-period-metallicity relation gives an estimate of the metallicity that is generally consistent with the spectroscopic method, with an uncertainty of 0.3 dex (see Fig. 9 in Kinemuchi et al. 2006).

\begin{center}
\setlength{\abovecaptionskip}{15pt plus 3pt minus 2pt}
\begin{figure}[th]
\includegraphics[width=0.45\textwidth]{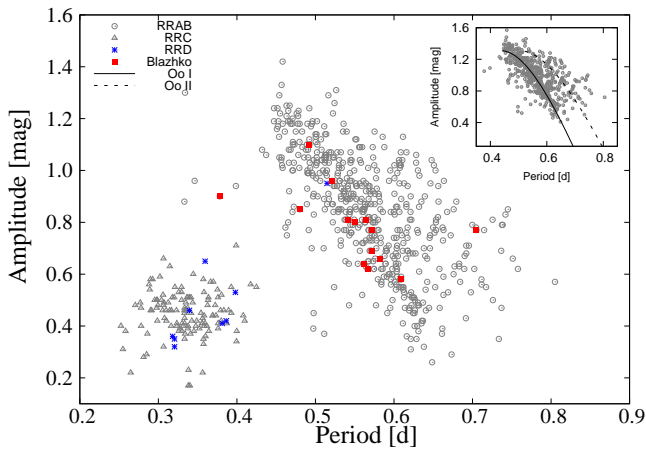}
\caption{The amplitude of light variation as a function of the period for the TNTS RR Lyrae stars. The circles and triangles represent the RRAB-type stars (fundamental overtone) and RRC-type stars(first overtone), respectively. The blue points and red points represent RRD-type stars and RRAB stars with Blazhko effect. The inset panel shows the Bailey diagram of the RRAB-type stars, with the
unfiltered amplitudes converted into the V-band values. The solid and dashed lines represent the locations of the Oosterhoff I and
Oosterhoff II subtypes (Zorotovic et al. 2010).}
\label{fig:amp}
\end{figure}
\end{center}

The distance to the RRAB Lyrae stars can be calculated using the relation given as:
\begin{equation}
d=10^{[(V_0)_s-M_V+5]/5}
\label{eq:distance}
\end{equation}
where $(V_0)_s$ represents the static average magnitudes in the V band. The Galactic extinctions of the TNTS unfiltered magnitudes were corrected from the recalibrated infrared dust maps (Schlafly \& Finkbeiner 2011). For the RR Lyrae sample in common, we found that the TNTS
unfiltered magnitudes have an offset of $\sim$0.1 mag relative to the V-band magnitudes from \citet[]{Drake2013}. This difference was
corrected in calculating the distances to the TNTS RRAB Lyrae stars. Assuming an averaged absolute magnitude $M_V$ = 0.6 mag for the
RRAB Lyrae stars \citep[]{Keller}, we can derive their heliocentric distances from the TNTS data. Then we converted these distances into
those relative to the Galactic plane of the Milkyway (i.e., $\vert$Z$\vert$).

Fig. 9 shows the spatial distribution of the metallicity derived for the TNTS RRAB Lyraes. The measurements from the CSS photometry and
the SDSS spectra are overplotted for comparison. As it can be seen, the photometric metallicities from the TNTS and CSS observations are
consistent with the spectroscopic estimates from the SDSS sample. The mean metallicity from the TNTS, the CSS and the SDSS is -1.59 dex,
-1.57 dex, and -1.64 dex, respectively. The inset of Figure 9 shows the TNTS RRAB Lyraes distinguished by the Oosterhoff types. From
Figures 8 and 9, one can see that the OoI population can occur at larger distances perpendicular to the Galactic plane
(i.e.,$\vert$Z$\vert$$\sim$45 kpc), and they have higher mean metallicity (i.e., [Fe/H] $\sim$ $-$1.5 dex) than the OoII population
(i.e, [Fe/H] $\sim$ $-$2.1 dex). The metallicity of RRAB Lyraes tends to have narrower distribution along the vertical distances away
from the Galactic plane. This tendency can be explained in part with that the RRAB Lyraes in OoI clusters dominate at larger vertical
distances, while a considerable fraction of these stars come from the OoII clusters (i.e., $\sim$22\% for the TNTS data) locating at
relatively smaller vertical distances (i.e., $\vert$Z$\vert$ $\leq$ 20 kpc). This dichotomy in metallcity distribution/Oosterhoff
classification of RRAB Lyrae stars suggests that the Galactic stellar halo was perhaps formed by two distinct populations, favoring the
dual-halo models (i.e., Miceli et al. 2008, An et al. 2013).

\begin{center}
\setlength{\abovecaptionskip}{15pt plus 3pt minus 2pt}
\begin{figure}[th]
\includegraphics[width=0.45\textwidth]{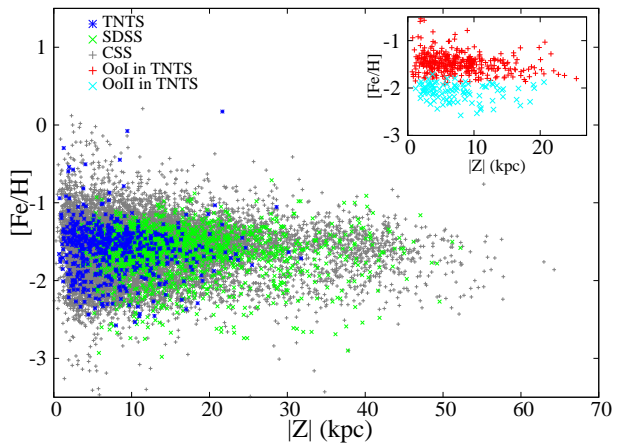}
\caption{The distribution of the metallicity measured for the RRAB stars versus their vertical distances from the galactic plane of the
Milkyway. The blue, green, and grey points represent the TNTS, SDSS, and CSS data, respectively. The red and blue points in
the inset panel represent the OoI and OoII subsamples of RRAB Lyraes from the TNTS.}
\label{fig:met-z}
\end{figure}
\end{center}

\subsubsection{Eclipsing Binary Stars}
Based on the variations of light curves, the eclipsing binaries can be classified into three broad categories: contact (EC), semidetached (ES), and detached (ED) \citep[]{Eclipse}. The EC binary represents a typical W UMa system (EW), which shows a continuous change in brightness and has an almost equal or un-obviously varying depth between the primary and secondary minima. The ES binary represents the Beta Lyrae-type system (EB), which shows a shallower change in brightness during the phase out of eclipse than the EC binary and has deeper primary depth than that of the secondary. The ED binary represents Algol-type system (EA), which shows almost constant brightness during the phase out of eclipse. Among our sample, there are 313 ECs, 88 ESs, and 30 EDs. For the 299 new variables, one third belongs to the eclipsing binary stars.

There are 67 interesting EC or ES binaries which exhibit the O'Connell effect that the two successive out-of-eclipse maxima are unequally high in the light curves \citep{Oconne51,Nataf10}. Models such as captured circumstellar material \citep{Liu03} and starspot activity \citep{Bell90} have been proposed to explain the O'Connell effect, but these explanations are still controversial. More photometric and spectroscopic observations are needed to understand the origin of such an effect seen in the eclipse binaries.

\subsubsection{Other Types of Variables}
The long periodic variable stars include Miras, semiregular pulsators, and slow irregular variables. Mira variable shows large-amplitude variabilities (i.e., $>$2.5 mag in $V$ band) with periods up to several hundred days. Semiregular pulsators show diverse periodicities in a range from several days to several thousand days; and the amplitude in light variations can vary from 0.01 mag to several magnitudes \footnote{http://www.sai.msu.su/gcvs/gcvs/iii/vartype.txt}. In our sample, 29 are Mira stars, 62 are semiregular pulsators, and 11 are slow irregular variables.

Observationally, Gamma Doradus stars show multiple periods ranging from 0.3 d to 3 d with a typical variation amplitude of less than 0.1 mag \citep{Cuypers2009}. Based on our determinations of the periodicity, 5 new ones (S070-6584, S598-3499, S619-728, SC49-6456, and SE61-2890) from our discovery perhaps belong to this class. Delta Scuti variables are late A- or early F-type pulsating stars. Their pulsation periods are found to be in the range of 0.02-0.25 days \citep{Breger2000}, such as star S585-17826 listed in Table \ref{tab:var}. W Virginis (CWA) is type II Cepheid which is regarded as the low mass analogues of the Cepheid \citep[]{Book}. SK39-3665 may be such a star in our sample.

\ \par
\ \par

\section{Conclusion}
We have presented the analysis of the unfiltered CCD images from the Tsinghua-NAOC transient survey. A total of 600 sky fields ($\sim$1300 square degrees), with repeated observations larger than 40 times during the period from October 2012 to August 2014, were selected for the detections of variable stars. We carried out photometry using SExtractor. Fluxes of all other images are normalized to the reference ones; and the resulting unfiltered magnitudes were further calibrated using the R-band magnitudes of the PPMX catalogue. We used SRPAstro, VARTOOLS and PDM to create light curves and detect variable stars.

We totally detected 1237 variables (including 299 new ones) with magnitude variations of $\gtrsim$ 0.1 mag and periods ranging from 0.1d to 500d. These variables include RR Lyraes, eclipsing binaries, and semiregular stars etc.. Among them 10 RRAB stars show the Blazhko effect and 64 eclipse binaries show the O'Connell effect.

We further examined the spacial distribution of the photometric metallicity for the RRAB stars and found that it shows a larger scatter near the Galactic plane (i.e., $-$3 $<$ [Fe/H] $<$ 0) but tends to converge at larger distances (i.e., $-$2 $<$ [Fe/H] $<$ $-$1. This variation can be explained with that the OoI type of RRAB Lyrae stars have a higher metallicity and can extend out to a larger vertical distance (i.e., $\vert$Z$\vert$ $\sim$ 45 kpc) while the OoII type of RRAB stars have a lower metallicity and they are relatively abundant at a vertical distance with $\vert$Z$\vert$ $\leq$ 20 kpc. This result favors that the Galactic halo was formed by two components with distinct properties.

\acknowledgements
We thank the anonymous referee for his/her insightful suggestions and Xiaobin Zhang for helpful discussions. This work was partly supported by the National Basic Research Program (973 Program) of China (Grant No. 2013CB834903 for Xiaofeng Wang, Xuefeng Wu, Grant Nos. 2012CB821800 and 2013CB834901 for Lingzhi Wang), the National Natural Science Foundation of China (NSFC grants 11178003, 11303041, 11322328, and 11325313), and  the Strategic Priority Research Program ``The Emergence of Cosmological Structures " (Grant No. XDB09000000) of the Chinese Academy of Sciences. Xiaofeng Wang is also supported by the Foundation of Tsinghua University (2011Z02170). Lingzhi Wang is also supported by the One-Hundred-Talent program of the Chinese Academy of Sciences (034031001). Xuefeng Wu is also supported by the One-Hundred-Talent Program and the Youth Innovation Promotion Association of the Chinese Academy of Sciences, and the Natural Science Foundation of Jiangsu Province.
\ \par

\ \par

\ \par

\bibliography{ms}{}
\bibliographystyle{apj}

\begin{sidewaystable}
\addtocounter{table}{1}
\LongTables
\begin{deluxetable*}{llllcrcrrrrcrl}
\tablenum{1}
\tablewidth{0pt}
\tablecaption{VARIABLE STARS}
\label{tab:var}
\tablehead{\colhead{(1)} & \colhead{(2)} & \colhead{(3)} & \colhead{(4)} & \colhead{(5)} & \colhead{(6)} & \colhead{(7)} & \colhead{(8)} & \colhead{(9)} & \colhead{(10)} & \colhead{(11)} & \colhead{(12)} & \colhead{(13)} & \colhead{(14)}\\
\colhead{Field} & \colhead{ID} & \colhead{R.A.} & \colhead{Dec.} & \colhead{Unfilter} & \colhead{Av} & \colhead{Amplitude} & \colhead{$J_s$} & \colhead{Period(TNTS)} & \colhead{Src$^{b}$}  & \colhead{Period(VSX)} & \colhead{$T_0^{c}$} & \colhead{Type$^{d}$} & \colhead{Note$^{e}$}\\
\colhead{     } & \colhead{  } & \multicolumn{2}{c}{(J2000.0)$^{a}$} & \multicolumn{3}{c}{(mag)} & \colhead{ } & \colhead{(d)} & \colhead{  } & \colhead{(d)} & \colhead{($HJD$-2456000)} & \colhead{    } & \colhead{    }}
\startdata
S007 & 424 & 22:47:55.16 & +25:46:59.60 & 13.88 & 0.19  & 0.33 & 0.89 & 0.382994 & (LS) & 0.382989 & 196.824 & EC & V,C \\
S008 & 1839 & 22:51:42.89 & +25:32:21.01 & 14.59 & 0.35  & 0.45 & 1.59 & 0.354250 & (LS) & 0.354265 & 196.650 & RRC & V,C \\
S008 & 542 & 22:53:00.98 & +25:43:48.03 & 14.84 & 0.34  & 0.23 & 1.23 & 46.094477 & (LS) &      & 214.880 & SR & N \\
S008 & 729 & 22:53:18.25 & +25:42:40.00 & 14.35 & 0.35  & 1.11 & 8.48 & 0.471269 & (LS) & 0.471272 & 196.318 & RRAB & V \\
S008 & 8407 & 22:54:12.33 & +24:38:07.03 & 15.41 & 0.26  & 1.02 & 2.46 & 0.608931 & (LS) & 0.608926 &      & RRAB & V,C \\
S008 & 8461 & 22:54:18.92 & +24:37:59.08 & 13.59 & 0.26  & 0.25 & 0.77 & 13.345449 & (LS) &      & 190.193 & SR & N \\
S009 & 9362 & 23:01:39.04 & +24:26:04.04 & 14.97 & 0.43  & 0.41 & 1.20 & 0.359450 & (PDM) & 0.359444 & 196.675 & RRC & V,C \\
S011 & 4138 & 23:11:27.60 & +25:08:42.01 & 17.18 & 0.18  & 1.04 & 0.85 & 0.546095 & (LS) & 0.546052 & 196.395 & RRAB & V,C \\
S012 & 9389 & 23:18:25.72 & +24:25:54.09 & 15.57 & 0.20  & 0.46 & 1.35 & 0.305437 & (LS) & 0.305435 & 196.489 & RRC & V,C \\
S012 & 9446 & 23:21:42.07 & +24:28:40.02 & 14.52 & 0.14  & 0.86 & 5.58 & 0.563126 & (LS) & 0.563094 & 196.381 & RRAB & V,C \\
S013 & 7367 & 23:24:04.12 & +24:32:33.09 & 16.61 & 0.11  & 1.00 & 1.57 & 0.509679 & (PDM) & 0.509708 & 196.495 & RRAB & V,C \\
S014 & 4043 & 23:35:53.93 & +25:08:56.05 & 16.41 & 0.14  & 1.00 & 1.03 & 0.549363 & (LS) & 0.549271 & 196.643 & RRAB & V,C \\
S014 & 6689 & 23:32:02.64 & +24:38:44.07 & 14.74 & 0.21  & 1.83 & 11.25 &      &    & 165.000000 & 229.217 & L & V \\
S014 & 6967 & 23:32:26.03 & +24:38:28.04 & 10.47 & 0.21  & 2.00 & 0.97 &      &    & 194.000000 &      & L & V \\
S015 & 7103 & 23:40:47.85 & +24:25:50.06 & 17.01 & 0.12  & 1.23 & 1.26 & 0.478438 & (PDM) & 0.478459 &      & RRAB & V,C \\
S037 & 10947 & 22:45:05.01 & +26:04:52.02 & 15.31 & 0.17  & 0.39 & 1.33 & 0.497227 & (LS) & 0.331645 & 196.565 & RRAB & V,C \\
S037 & 8742 & 22:43:33.57 & +26:18:47.05 & 17.93 & 0.14  & 1.31 & 0.59 & 0.526378 & (LS) & 0.526361 & 196.609 & RRAB & V,C \\
S038 & 10429 & 22:52:06.31 & +25:59:50.02 & 16.34 & 0.37  & 1.18 & 1.13 & 0.489781 & (PDM) & 0.489789 & 196.636 & RRAB & V,C \\
S038 & 12180 & 22:53:18.19 & +25:42:47.00 & 14.44 & 0.35  & 1.10 & 4.87 & 0.471307 & (LS) & 0.471272 & 196.295 & RRAB & V \\
S039 & 10942 & 22:59:08.64 & +25:49:50.09 & 14.78 & 0.36  & 0.90 & 5.48 & 0.497052 & (LS) &      & 196.866 & RRAB & N

\enddata
\tablecomments{\\
  $^a$: Based on the TNTS reference images. \\
  $^b$: L-S: Lomb-Scargle method; PDM: Phase Dispersion Minimization method; VSX: from VSX catalogs. \\
  $^c$: Epoch when primary eclipse or minimum light occurs. \\
  $^d$: ACEP= Anomalous Cepheids; RS= RS Canum Venaticorum; EC= contact binary; ES= semi-detached binary; ED= detached binary; CWA= W Virginis; RR= RR~Lyrae variable; BL= ``Blazhko effect''; CEP= Cepheids ; GDOR= Gamma Doradus star; DSCT= Delta Scuti star; SR= Semiregular pulsator; M= Mira variable; LPV= long period variable; L= slow irregular variable \\
  $^e$: V= AAVSO VSX variable; C= CSS variable; L= LINEAR variable; N= New variable.}

\end{deluxetable*}
\end{sidewaystable}

\begin{appendix}
\newcounter{subfigure}
\setcounter{subfigure}{1}
\renewcommand{\thefigure}{\arabic{figure}\alph{subfigure}}

\begin{figure}[t]
\begin{center}
\includegraphics[width=1\textwidth]{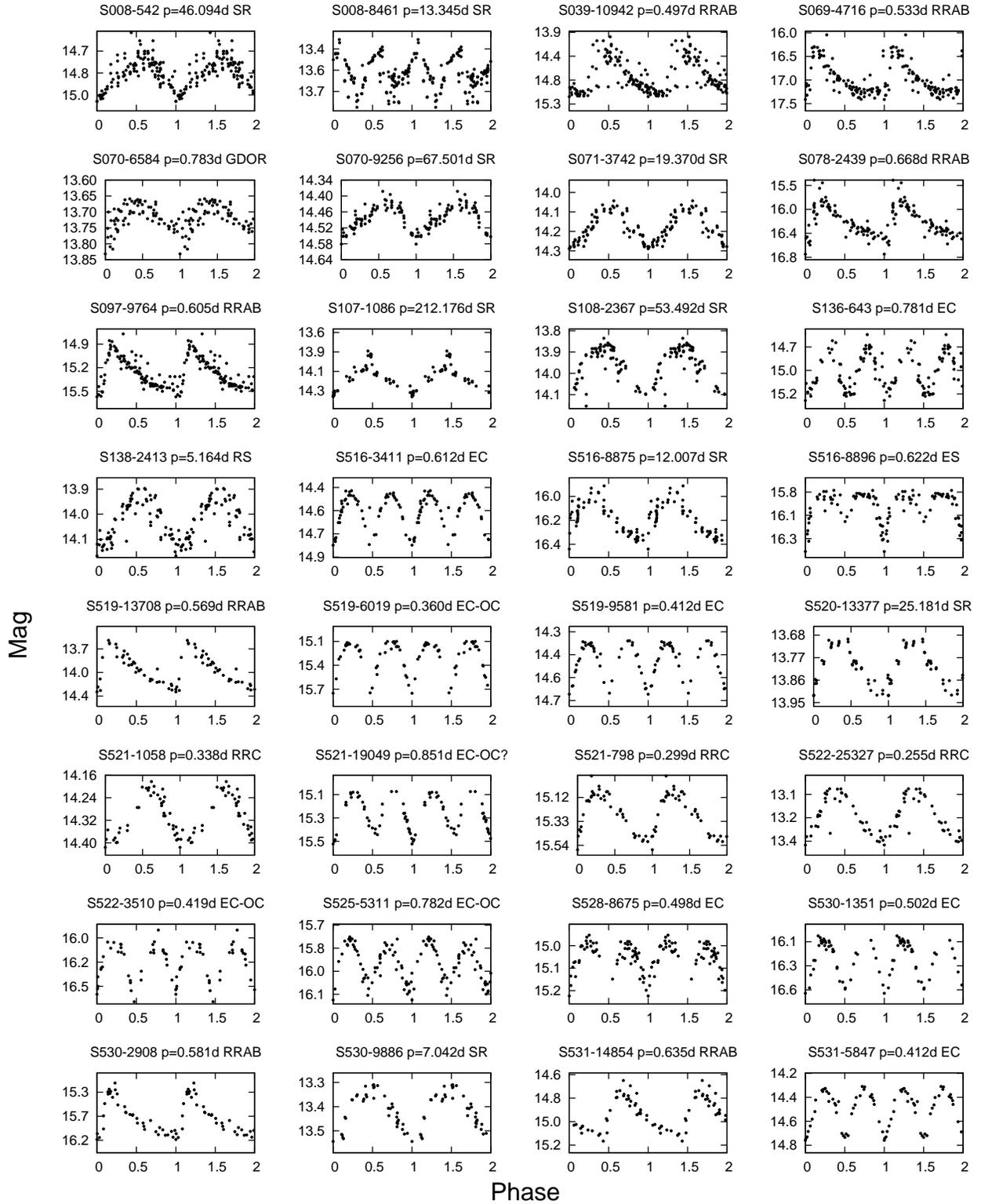}
\caption{Phase diagrams of new variables. }
\label{fig:}
\end{center}
\end{figure}

\begin{figure}[t]
\begin{center}
\includegraphics[width=0.9\textwidth]{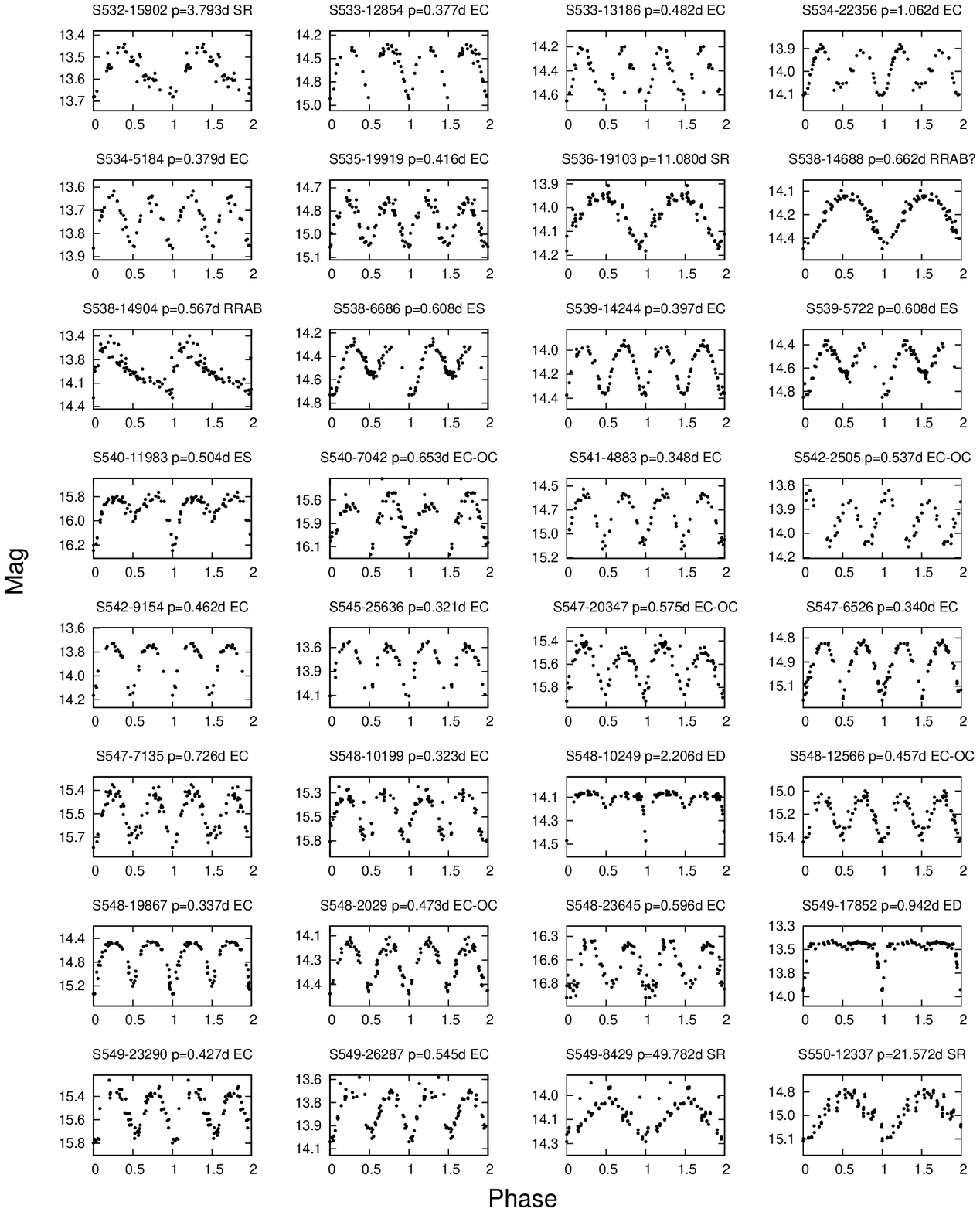}
\caption{Continued. }
\end{center}
\end{figure}

\begin{figure}[t]
\begin{center}
\includegraphics[width=1\textwidth]{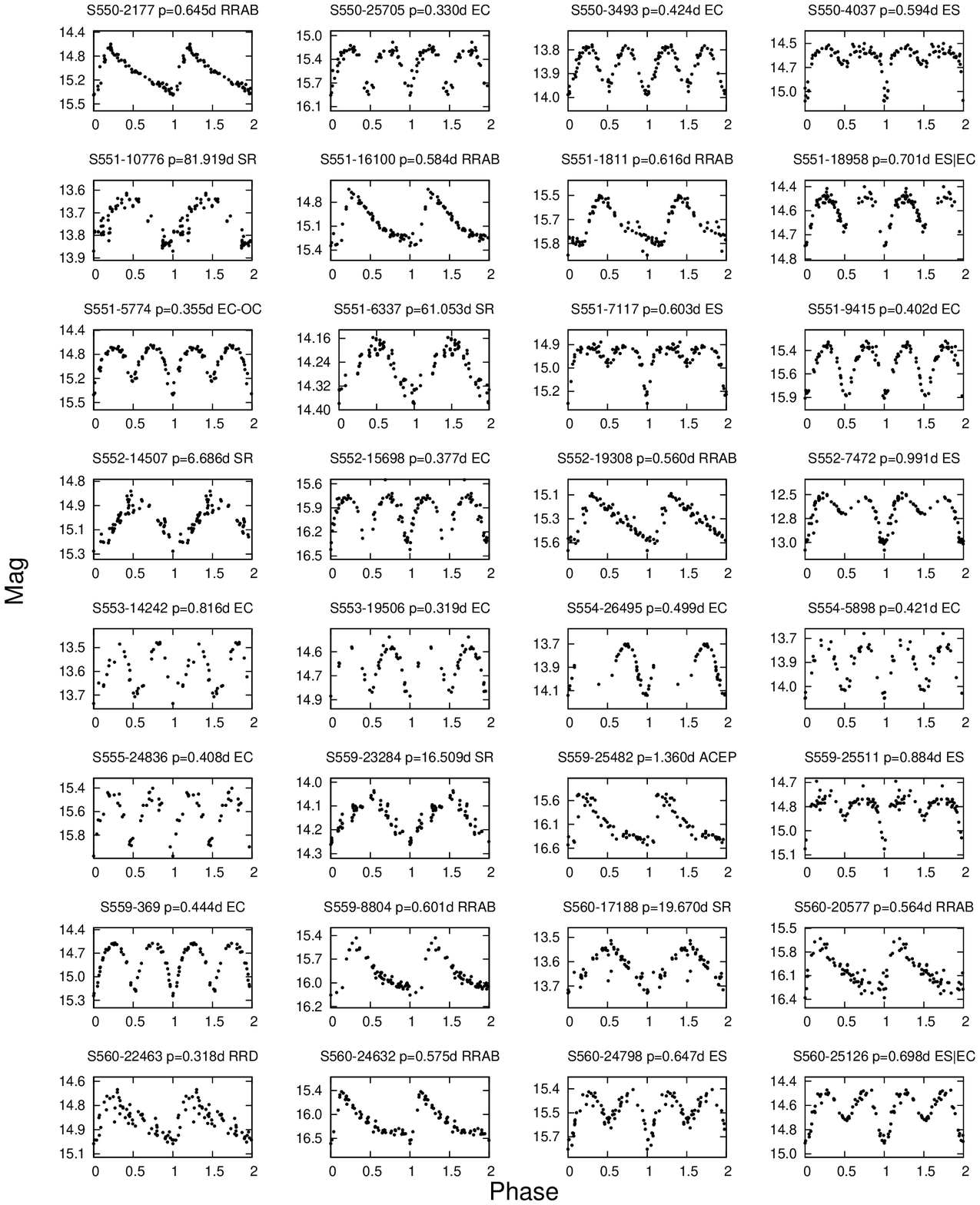}
\caption{Continued. }
\label{fig:ecl1}
\end{center}
\end{figure}

\begin{figure}[t]
\begin{center}
\includegraphics[width=0.9\textwidth]{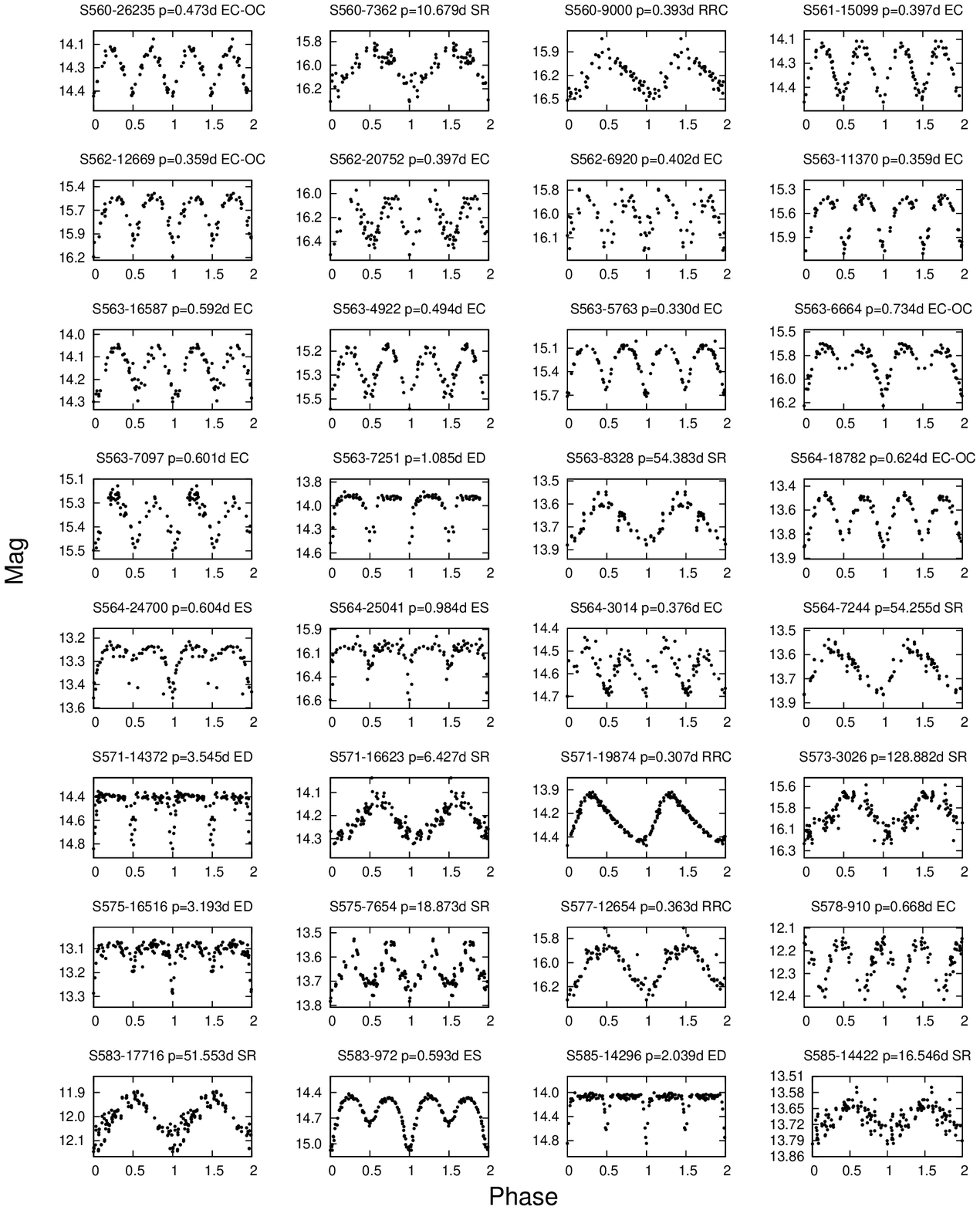}
\caption{Continued. }
\end{center}
\end{figure}

\begin{figure}[t]
\begin{center}
\includegraphics[width=0.9\textwidth]{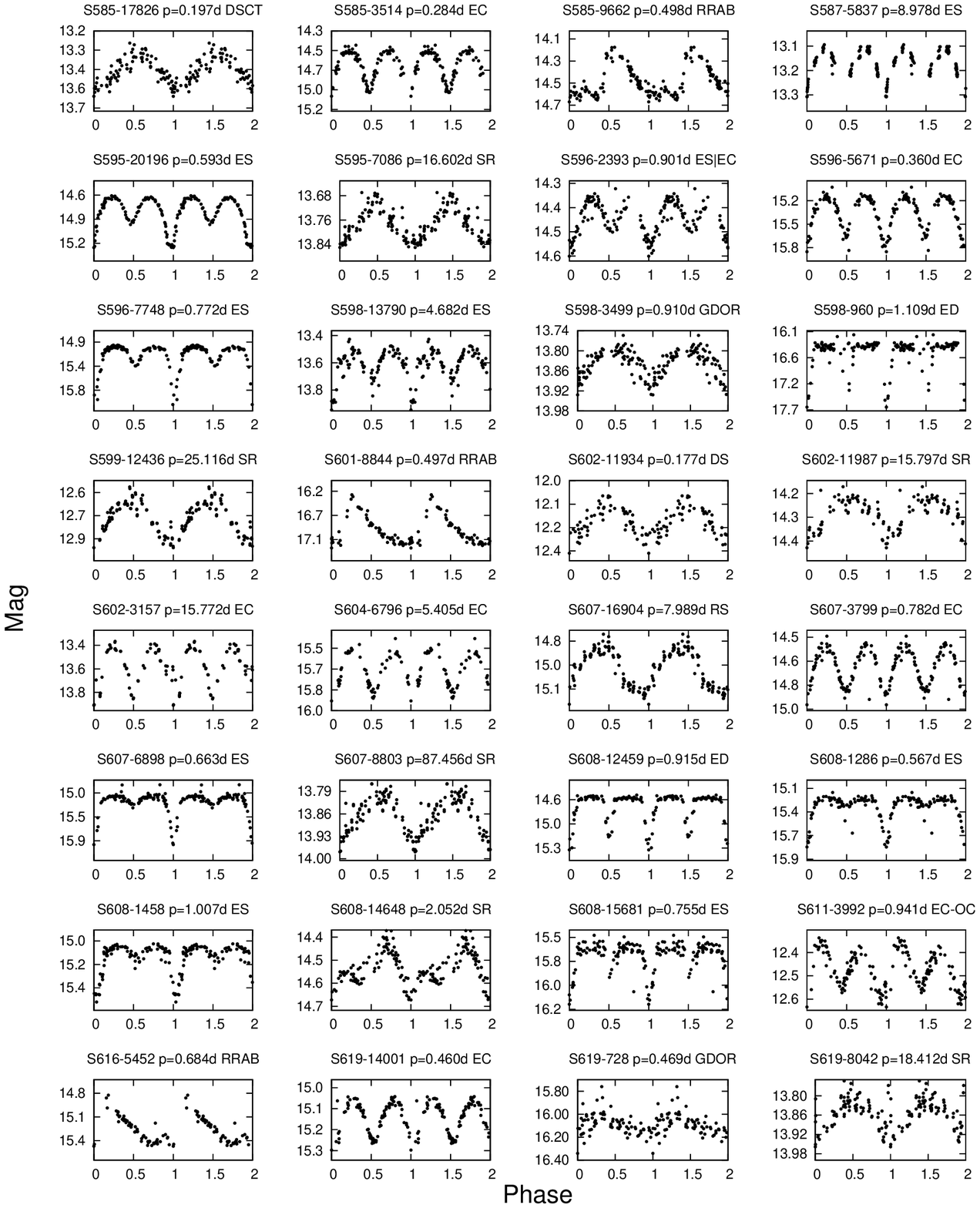}
\caption{Continued. }
\end{center}
\end{figure}

\begin{figure}[t]
\begin{center}
\includegraphics[width=1\textwidth]{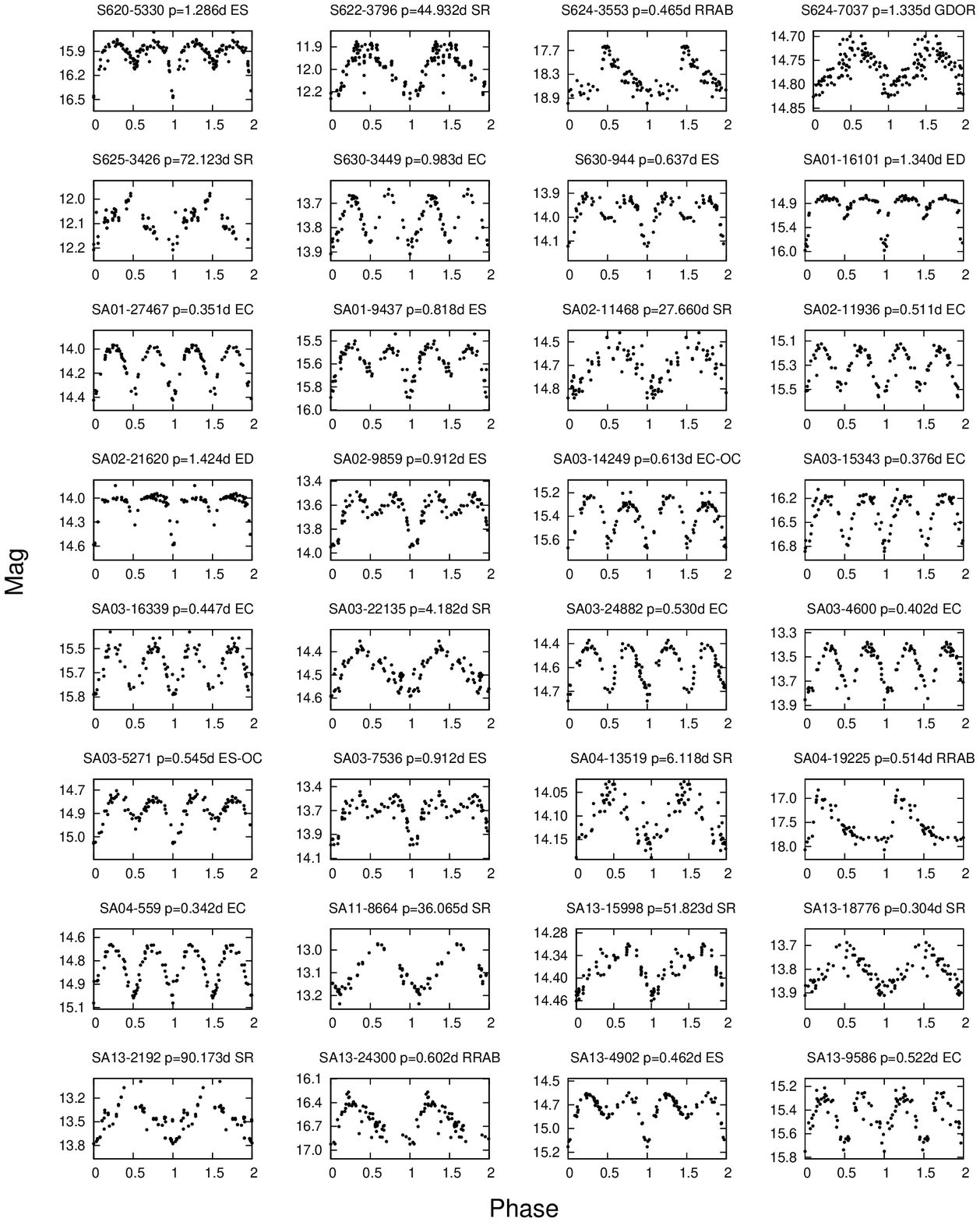}
\caption{Continued. }
\label{fig:ecl1}
\end{center}
\end{figure}

\begin{figure}[t]
\begin{center}
\includegraphics[width=0.9\textwidth]{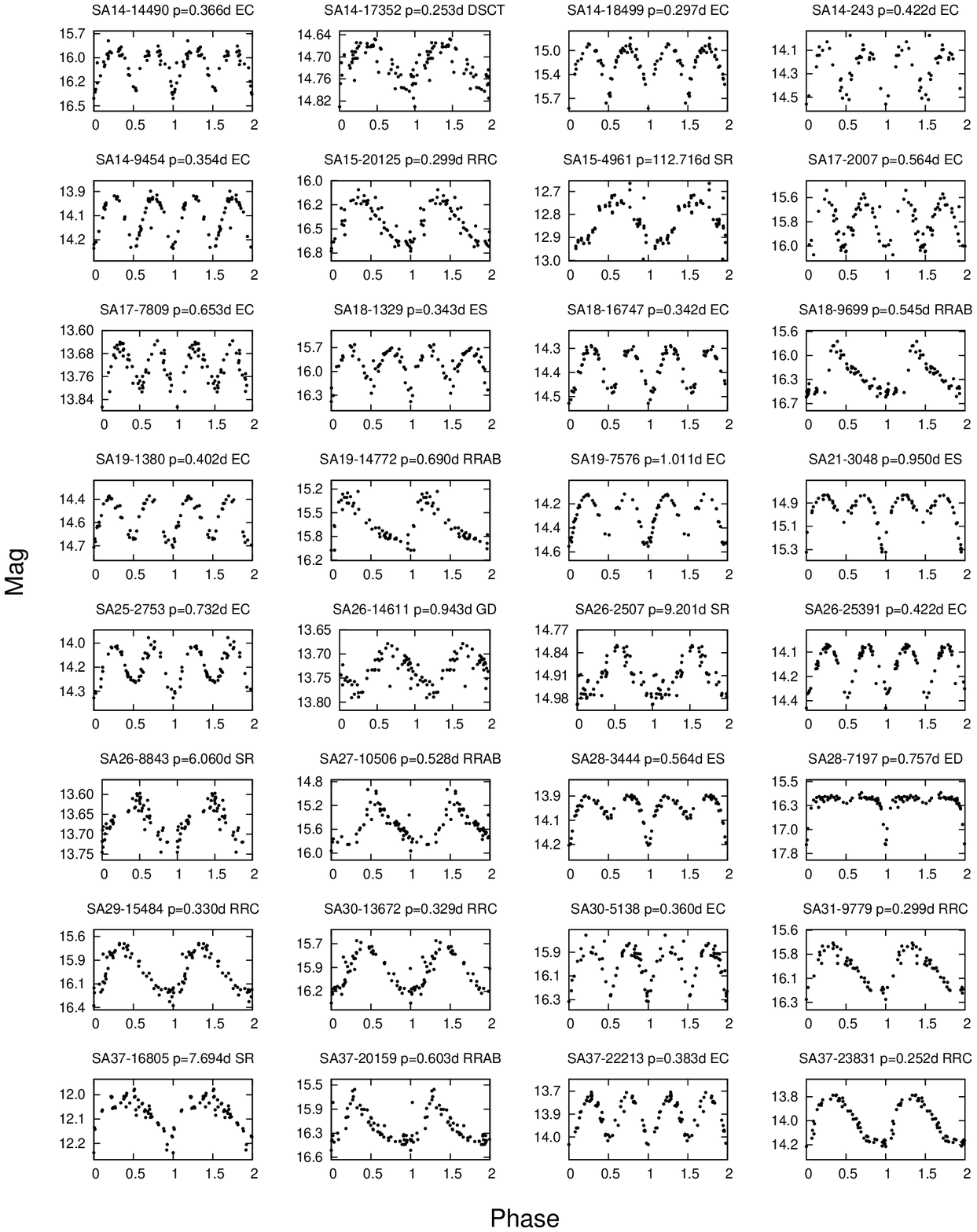}
\caption{Continued. }
\end{center}
\end{figure}

\begin{figure}[t]
\begin{center}
\includegraphics[width=0.9\textwidth]{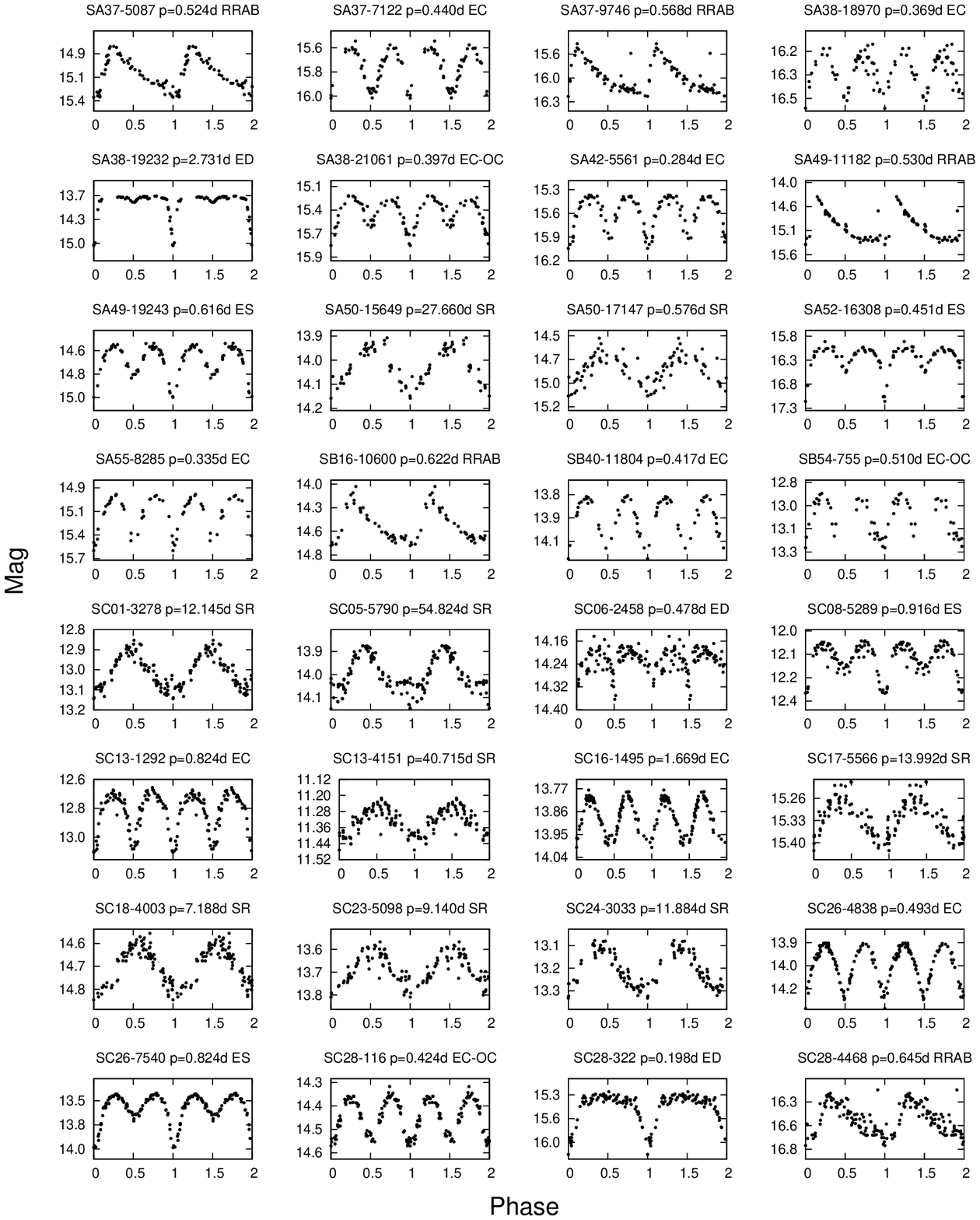}
\caption{Continued. }
\end{center}
\end{figure}

\begin{figure}[t]
\begin{center}
\includegraphics[width=1\textwidth]{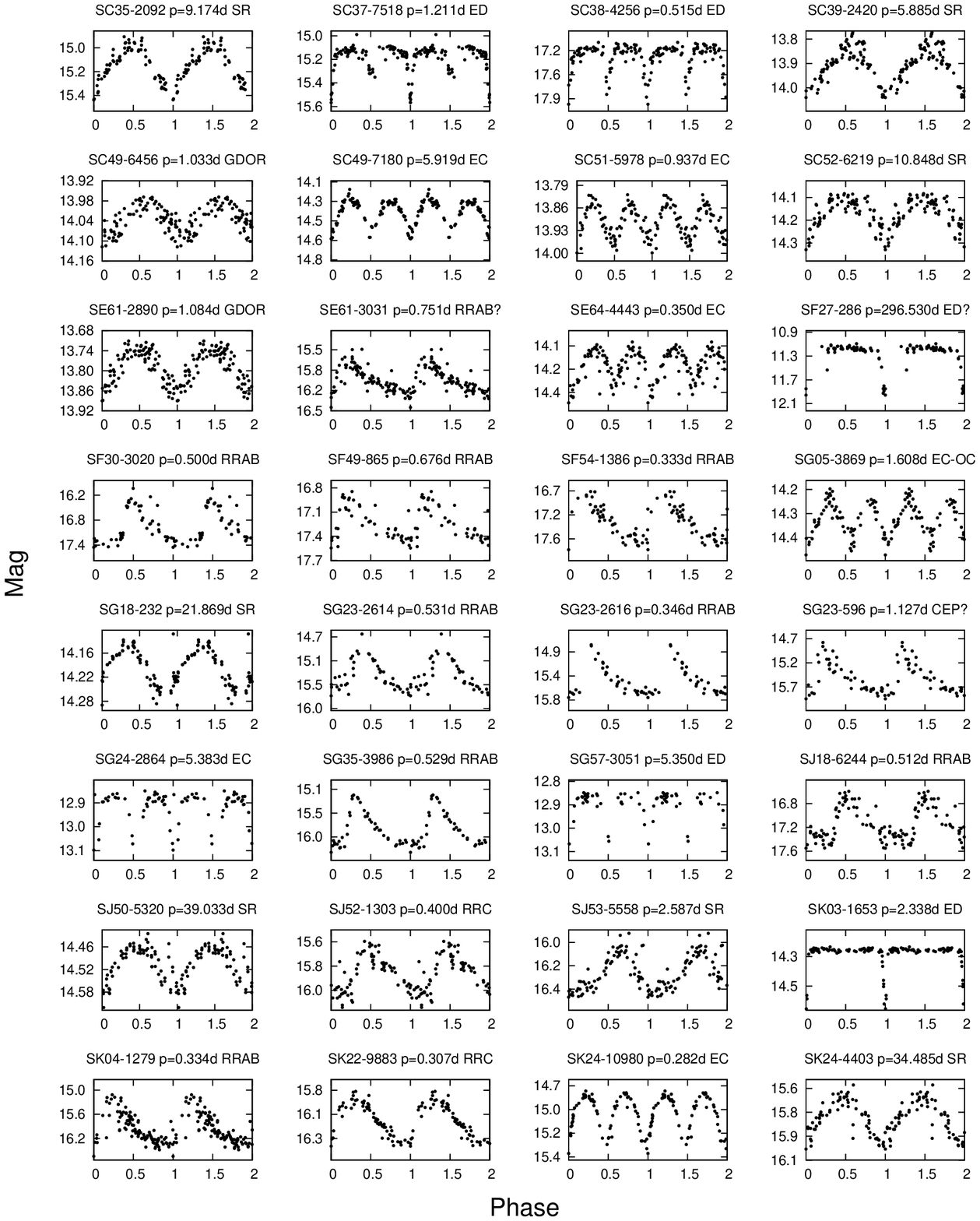}
\caption{Continued. }
\label{fig:ecl1}
\end{center}
\end{figure}

\begin{figure}[t]
\begin{center}
\includegraphics[width=0.9\textwidth]{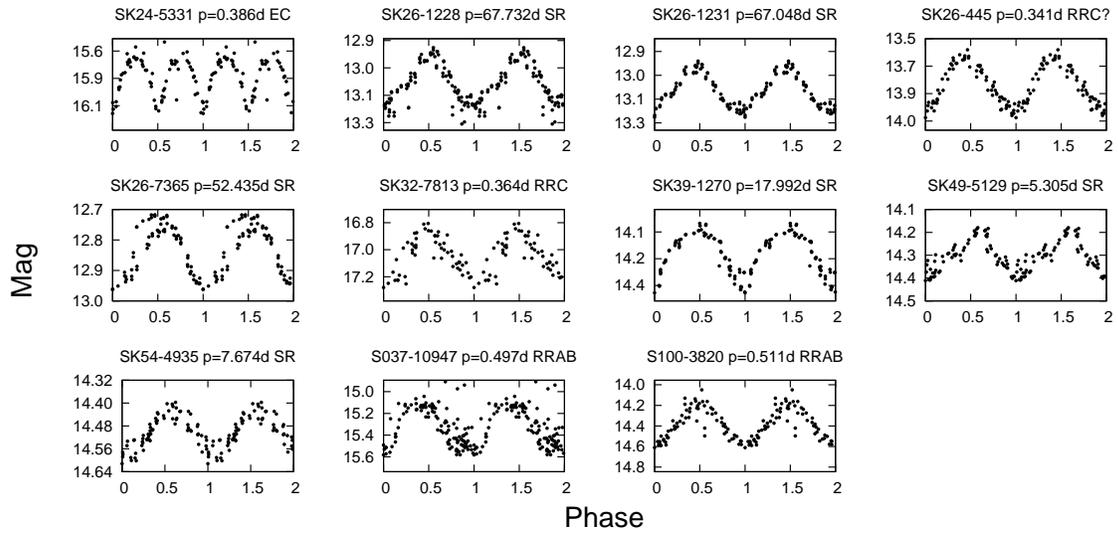}
\caption{Continued (Including two special RRAB stars which are not new). }
\end{center}
\end{figure}

\end{appendix}
\end{document}